\documentclass[aps,preprint]{revtex4}
\usepackage{graphicx}
\usepackage{dcolumn}
\usepackage{bm}
\usepackage{color}
\usepackage{amsmath}

\newcommand{\ket}[1]{\left|{#1}\right\rangle}

\begin{document}

\title{Slow light using spectral hole burning in a Tm${}^{3+}$:YAG crystal}

\author{R. Lauro}
\email{romain.lauro@lac.u-psud.fr}
\author{T. Chaneli\`{e}re}
\author{J. L. Le Gou\"{e}t}

\affiliation{Laboratoire Aim\'{e} Cotton, CNRS-UPR 3321, Univ Paris-Sud, B\^{a}t. 505, 91405 Orsay cedex, France.}

\begin{abstract} We report on light slowing down in a rare earth ion doped crystal by persistent spectral hole burning. The absence of motion of the active ions, the large inhomogeneous broadening, the small homogeneous width and the long lifetime of the hyperfine shelving states make this material convenient for the burning of narrow persistent spectral holes. Since the hole can be burnt long before the arrival of the input signal, there is no need for a strong coupling field, illuminating the sample simultaneously with the input signal, in contrast with procedures such as Electromagnetically Induced Transparency or Coherent Population Oscillations. Analyzing the slowing down process, we point out the role played by off resonance atoms where most of the incoming information is carried over while the pulse is confined within the sample.\end{abstract}

\maketitle

\section{Introduction}
A light pulse can propagate in a resonant medium faster or slower than in vacuum. This is a consequence of the wave nature of light~\cite{brillouin1960}. Studying light velocity has some fundamental interests~\cite{garrett1970,chu1982,wang2000}, and may have some practical applications in non linear optics~\cite{harris1990} and optical information processing~\cite{camacho2007}. Moreover, slow light has been brought into play in quantum storage investigations where a quantum state of light is mapped into an atomic ensemble~\cite{lukin2003}. Since the observation of a group velocity of $17\;m{s}^{-1}$ in an ultracold atomic gas~\cite{hau1999}, slow light has become an intense field of research~\cite{kasapi1995,wei1999,budker2000,turukhin2002,bigelow2003,longdell2005,baldit2005}, making use of several techniques, in different media.

Most of slow light experiments used the Electromagnetically Induced Transparency phenomenon (EIT)~\cite{harris1997} in cold atoms~\cite{hau1999}, warm vapors~\cite{kasapi1995}, as well as in rare earth ions doped solids~\cite{turukhin2002,longdell2005} and nitrogen vacancy centers in diamond ~\cite{wei1999}.
Another phenomenon leading to delayed light are the Coherent Population Oscillations (CPO)~\cite{shwarz1967}, with experimental realization in ruby~\cite{bigelow2003} and erbium doped crystal~\cite{baldit2005}. However, in EIT and CPO protocols, a strong coupling field has to be shined to the atoms together with the signal pulse. The intense coupling field may hamper the monitoring and detection of the weak probe.

Recently, it has been proposed to use spectral hole burning for reducing the speed of light~\cite{shakhmuratov2005,rebane2007}. So far, the only relevant experimental demonstration was achieved in hot rubidium vapor~\cite{camacho2006}. To the best of authors' knowledge, there is no such demonstration in Rare Earth Ions doped Crystals (REIC), while there are some strong advantages with this material. First, atoms are motionless, hence the hole lifetime is not limited by atomic diffusion. One the other hand, due to the long hyperfine state lifetime (several seconds for Tm${}^{3+}$:YAG), it is possible to efficiently pump atoms in a non-resonant state for a long time: the signal pulse can be sent after hole was burnt, contrary to Ref.~\cite{camacho2006}, where pump and signal fields are simultaneous. Moreover, with an optical dipole lifetime of tens of microseconds, typically four orders of magnitude larger than in alcalines atoms, one can burn holes as narrow as a few hundreds of kHz. Finally, the large ratio between inhomogeneous and homogeneous broadenings available in REIC offers possibilities for multimode delay lines.

This paper is organized as follows. We first give a theoretical description of slow light propagation in hole burning medium, and show relationship with adiabatic following. Then we show some experimental results in a Tm${}^{3+}$:YAG crystal.

\section{Slow light propagation in a spectral hole: atomic response and energy transfer}
An electromagnetic signal of frequency $\omega_{0}$ propagating in a dispersive medium has a group velocity c/$[n+\omega_{0} (dn/d \omega)]$, where n($\omega$) is the refractive index. Slow light is obtained with strong normal dispersion ($dn/d \omega>>1/\omega_{0}$). As shown by the Kramers Kronig relations, it is possible to obtain such a strong positive slope for refractive index if a transparency window (TW) is created in an absorbing medium. Deeper and narrower the TW, slower the light pulse. The group velocity $v{}_{g}$ is of the order $\Delta_{0} / \alpha_{0}$, where $\Delta_{0}$ and $\alpha_{0}$ respectively stand for the TW width and the absorption coefficient outside TW. Such a structure is commonly created with hole burning spectroscopy in inhomogeneously broadened materials. An engraving laser pulse modifies thermalized atomic populations over a spectral range, larger than the homogeneous width, but narrower than the inhomogeneous one. It is possible to achieve spectral hole burning in a two-level-system to slow down light~\cite{shakhmuratov2005,rebane2007}: in this case, linear absorption coefficient depends on the atomic population difference, and transparency is obtained if both ground and excited populations are equal. However, the spectral hole lifetime is limited by the excited state lifetime. In three-level $\Lambda$-like atoms, optical pumping to a shelving state may lead to a much longer hole lifetime. In REIC, such $\Lambda$-systems can be built on the ground state hyperfine structure, with a corresponding lifetime of up to a few days~\cite{konz2003}.

An incoming signal pulse, narrower than the previously burnt spectral hole, shall exit the medium with little shape distortion and energy loss. However, during its propagation through the sample at velocity $v{}_{g}$, the signal is spatially compressed by a factor $v{}_{g}$/c. If the medium is thick enough to accomodate the entire compressed pulse, the optically carried energy is reduced by the same factor. If $v{}_{g}$$<<$c, most of the electromagnetic energy escapes from the pulse as it travels through the medium, but is recovered at the sample exit. This energy must be temporarily stored in the material.

Relationship between slowing down and energy transfer to atoms was explained in terms of adiabatic following by Grischkowsky~\cite{grischkowsky1973} in the case of an optical pulse tuned to the wing of an absorption line in rubidium vapor. In Ref.~\cite{grischkowsky1973}, since the optical detuning is much larger than inhomogeneous linewidth, atoms are all excited in the same way, and only homogeneous dephasing time plays a role. Here, we extend the work of Grischkowsky to the case of spectral hole burning, taking into account inhomogeneous broadening. We consider an ensemble of two level (ground state $\left|1\right\rangle$ and excited state $\left|2\right\rangle$) atoms with optical frequency $\omega_{12}$ excited by a weak signal field $E(z,t) = \frac{1}{2}\mathcal{E}(z,t)e{}^{i(kz-\omega_{l}t)}$+c.c. (k=$\omega_{l}$/c). Within the framework of the rotating wave approximation, the optical Bloch equations read as:
\begin{eqnarray}
\dot{\widetilde{\sigma}}_{21}(z,t) = \frac{i}{2} \Omega(z,t)w(z,t) - (i\Delta + \frac{1}{T_{2}})\sigma_{21}(z,t)\label{coherence}
\end{eqnarray}
\begin{eqnarray}
\dot{w}(z,t) = - i \Omega(z,t)(\widetilde{\sigma}_{12}(z,t) - \widetilde{\sigma}_{21}(z,t)) + \frac{1}{T_{1}}(1 - w(z,t))\label{population}
\end{eqnarray}
where $\sigma_{ij}(z,t)=\widetilde{\sigma}_{ij}(z,t)e^{i(kz-\omega_{l}t)}$ is the optical coherence and $w(z,t)$ is the population difference between ground and excited states ($w=\sigma_{11}-\sigma_{22}$), $\Omega(z,t)=\mu_{12}\ \mathcal{E}(z,t)/\hbar$ is the Rabi frequency ($\mu_{12}$ is the dipole moment matrix element), T${}_{1}$ and T${}_{2}$ are the longitudinal and transverse relaxation times, respectively, $\Delta$ = $\omega_{12}$ - $\omega_{l}$ is the optical detuning. Since the probe pulse is weak, $w(z,t)\approx1$. We can write optical coherence as follows:
\begin{eqnarray}
\widetilde{\sigma}_{21}(z,t) = \frac{i}{2} e^{-(i\Delta + 1/T_{2})t} \int_{-\infty}^{t} \Omega(z,t^{'}) e^{(i\Delta + 1/T_{2})t^{'}} dt^{'}\label{coherence_int1}
\end{eqnarray}
With the condition $\lim_{t \rightarrow -\infty}\Omega(t)=0$, repeated integrations by parts lead to:
\begin{eqnarray}
\widetilde{\sigma}_{21}(z,t) = -\frac{i}{2} \sum_{n=0}^{\infty} \left(\frac{-1}{i \Delta + 1/T_{2}}\right)^{n+1} \frac{\partial^{n}}{\partial t^{n}}  \Omega(z,t)\label{coherence_int2}
\end{eqnarray}
We express the macroscopic density of polarization in terms of the atomic coherence as $P(z,t) = \mu_{12} \int G(\omega_{12}) (\sigma_{12}(z,t) + \sigma_{21}(z,t))  d\omega_{12}$, where the atomic inhomogeneous distibution of width $\Gamma_{inh}$ is denoted $G(\omega_{12})$. 
\begin{figure}[!ht]
\begin{centering}
\includegraphics[width=8cm]{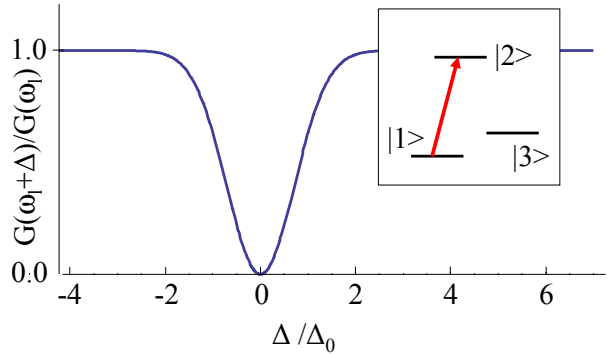}
\caption{Inhomogeneous distribution $G(\Delta+\omega_{l})/G(\omega_{l})$ with a $\Delta_{0}$-wide spectral hole. The inset represents the $\Lambda$-shape three-level system. The input signal pulse excites the transition 1-2}\label{hole}
\end{centering}
\end{figure}
When a spectral hole is burnt, this distribution is modified. For sake of simplicity, we assume that the spectral hole (centered at $\omega_{l}$, as well as G) has a lorentzian shape of width $\Delta_{0}$, much narrower than $\Gamma_{inh}$. The hole must be broader than the pulse spectrum, and the pulse duration must be shorter than the dephasing time $T_{2}$, which entails the condition $1/T_{2}<<\Delta_{0}<<\Gamma_{inh}$. The modified inhomogeneous distribution is defined by $G^{'}(\omega_{12})=G(\omega_{12})\left(1-\left(\Delta_{0}^{2}/4\right)/\left[ \left( \omega_{12}-\omega_{l} \right) ^{2}+\Delta_{0}^{2}/4 \right]\right)$. Hence, the positive frequency component of polarization can be written:
\begin{eqnarray}
\displaystyle \mathcal{P}(z,t) &= & - i \frac{\mu_{12}}{2} \sum_{n=0}^{\infty} \frac{\partial^{n}}{\partial t^{n}}\Omega(z,t)  \int_{-\infty}^{\infty} G(\Delta+\omega_{l}) \left(\frac{-1}{i \Delta + 1/T_{2}}\right)^{n+1} \left(1-\frac{\Delta_{0}^{2}/4}{\Delta^{2}+\Delta_{0}^{2}/4} \right) d \Delta \label{polar2}
\end{eqnarray}
where $\Delta=\omega_{12}-\omega_{l}$. In Integrating Eq.~\ref{polar2}, due to the spectral hole, most of the contribution comes from atoms with a detuning larger than $\Delta_{0}$. For these atoms, one can write: $\left(-1/\left[i \Delta + 1/T_{2}\right]\right)^{n+1}\approx\left(i/\Delta \right)^{n+1}$. For even $n$-index values, $\left(i/\Delta \right)^{n+1}$ is an odd function. The contributions from the atoms with $\Delta$ and $-\Delta$ detunings cancel each other, whereas for odd $n$-index values these contributions combine constructively. Moreover, considering that the signal pulse varies slowly on $1/\Delta_{0}$-timescale, we reduce Eq.~\ref{polar2} to the $n=1$ leading term, neglecting higher order derivatives. Keeping in mind that inhomogeneous broadening is much larger than $1/T_{2}$ and $\Delta_{0}$, one can write the macroscopic polarization as:
\begin{eqnarray}
\mathcal{P}(z,t) & = & \frac{i \mu_{12}^{2} G(\omega_{l})}{2 \hbar} \frac{\partial}{\partial t}\mathcal{E}(z,t) \int_{-\infty}^{\infty}  \left(\frac{1}{\Delta^{2}} \right) \left(1-\frac{\Delta_{0}^{2}/4}{\Delta^{2}+\Delta_{0}^{2}/4} \right) d\Delta \label{polar3}\nonumber \\
                 & = & i\frac{ \pi\mu_{12}^{2} G(\omega_{l})}{\hbar \Delta_{0}} \frac{\partial}{\partial t}\mathcal{E}(z,t)\label{polar_final}
\end{eqnarray}
Inserting $\mathcal{P}(z,t)$ into the linearized propagation equation
\begin{eqnarray}
\frac{\partial}{\partial z}\mathcal{E}(z,t) + \frac {1}{c}\frac{\partial}{\partial t}\mathcal{E}(z,t) = i \frac{k}{\epsilon_{0}}\mathcal{P}(z,t)\label{propagation}
\end{eqnarray}
leads to the travelling wave equation:
\begin{eqnarray}
\frac{\partial}{\partial z}\mathcal{E}(z,t) + \left(\frac {1}{c}+\frac{\alpha_{0}}{\Delta_{0}}\right)\frac{\partial}{\partial t}\mathcal{E}(z,t)=0 \label{propagation2}
\end{eqnarray}
where we have substituted the linear absorption coefficient $\alpha_{0} = \pi k G(\omega_{l})\mu_{12}^{2} / \hbar \epsilon_{0}$. The signal propagates without distortion at velocity $v_{g}$ given by:
\begin{equation}\label{group_velocity}
1/v_{g} = 1/c + \alpha_{0}/\Delta_{0}
\end{equation}
In REIC, typical values of $\alpha_{0}$ and $\Delta_{0}/2\pi$ are around 10${}^{3}$ m${}^{-1}$ and 100 kHz respectively, which gives a light velocity reduction factor $v_{g}/c\approx 3 . 10^{-6}$. At the input and output sides of the atomic medium the signal pulse amplitude is preserved since the index of refraction is unity. Hence, the signal pulse inside medium reads as:
\begin{eqnarray}
\mathcal{E}^{in}(z,t) = \mathcal{E}^{out}(z=0,t-z/v_{g}) \label{propagation3}
\end{eqnarray}
where $\mathcal{E}^{in}(z,t)$ and $\mathcal{E}^{out}(z,t)$ respectively stand for the electric field inside and outside medium.

We turn now to the energy transfer from light to atoms. From Eq.~\ref{population}, we can evaluate the excited population:
\begin{eqnarray}
\sigma_{22}(z,t) = \frac{i}{2} e^{-t/T_{1}} \int_{-\infty}^{t} \Omega(z,t^{'}) \left( \sigma_{12}(z,t) - \sigma_{21}(z,t) \right) e^{t^{'}/T_{1}} dt^{'} \label{pop_excited}
\end{eqnarray}
Both the pulse duration and propagation time are much smaller than T${}_{1}$, so Eq.~\ref{pop_excited} can be written:
\begin{eqnarray}
\sigma_{22}(z,t) = \int_{-\infty}^{t} \Omega(z,t^{'}) \mathrm{Im} \left( \sigma_{21}(z,t^{'}) \right) dt^{'} \label{pop_excited2}
\end{eqnarray}
Once again, we only have to consider far detuned atoms ($\Delta>>1/T_{2},\Delta_{0}$). In Eq.~\ref{coherence_int2}, the term corresponding to $n=0$ is real, so we can write: $\mathrm{Im} \left( \sigma_{21} \right) = \frac{\partial}{\partial t} \Omega /(2 \Delta^{2})$, which leads to the expression of the excited population:
\begin{eqnarray}
\sigma_{22}(z,t) = \frac{1}{2 \Delta^{2}} \int_{-\infty}^{t} \Omega(z,t^{'}) \frac{\partial}{\partial t^{'}} \Omega(z,t^{'}) dt^{'} = \frac{1}{4} \left(\frac{\Omega(z,t)}{\Delta} \right)^{2} \label{pop_excited3}
\end{eqnarray}
Integrating over the spatial and spectral atomic distributions, one obtains the energy stored in the atoms at time $t$  
\begin{eqnarray}\label{energy1}
W_{at} & = & \frac{\mu_{12}^{2} \omega_{12}}{4 \hbar} \int_{0}^{L} \left|\mathcal{E}^{in}(z,t)\right|^{2} dz \int_{-\infty}^{\infty} \frac{1}{\Delta^{2}} G(\Delta+\omega_{l}) \left(1-\frac{\Delta_{0}^{2}/4}{\Delta^{2}+\Delta_{0}^{2}/4} \right) d\Delta \nonumber\\
& = & c \frac{\alpha_{0}}{\Delta_{0}} \frac{\epsilon_{0}}{2} \int_{0}^{L} \left|\mathcal{E}^{in}(z,t)\right|^{2} dz 
\end{eqnarray}
In terms of optically carried energy inside the medium, $W_{em}^{in} = \epsilon_{0} / 2 \int_{0}^{L} \left|\mathcal{E}^{in}(z,t)\right|^{2} dz$, and of group velocity (see Eq. \ref{group_velocity}), Eq. \ref{energy1} can be expressed as:
\begin{equation}\label{energy_at}
W_{at}=\left(\frac{c}{v_{g}}-1\right)W_{em}^{in} 
\end{equation}
In slow light regime, most of energy is stored in the atoms.

In the above discussion both the slowing down process and the temporary energy transfer to off-resonance atoms are derived from the Maxwell-Bloch equations. At the end we verify the energy conservation. A different approach may be considered, where the speed of light reduction is derived from the conservation of energy. As the signal wave travels through the medium, the off-resonance atoms stay in the adiabatic state $c_+(z,t)\ket 1 +c_-(z,t)\ket 2$, where  $c_\pm(z,t)=(\sqrt{\Omega(z,t)^2+\Delta^2}\pm\Delta)^{1/2} /{\sqrt{2}(\Omega(z,t)^2+\Delta^2)^{1/4}}$. Expanding state $\ket 2$ population in powers of $ \Omega(z,t)/\Delta$, one recovers Eq.\ref{pop_excited3}. Summing the excited state energy over all the atoms leads to Eq. \ref{energy1}. The conservation of energy can be expressed as $W_{em}^{out}=W_{at}+W_{em}^{in}$. The energy carried by the signal pulse before entering the medium, $W_{em}^{out}$, is shared between the atoms and the electromagnetic field when the pulse is confined in the medium. Besides, $W_{em}^{in}=W_{em}^{out}v/c$, since the light pulse behaves as a travelling wave. Hence the group velocity inside the medium is finally given by $v/c=(W_{em}^{out}-W_{at})/W_{em}^{out}$.
        
Rather paradoxically, the total energy travels at group velocity in the medium, as confirmed by the light pulse revival at the exit of the sample, although the electromagnetic field carries nearly no energy inside the medium. This is rather surprising since the energy can only be transported by this weak field. The consistency can only be restored if the optically carried energy travels at speed $c$. Indeed, let $U_{em}^{in}$ (respectively $U_{at}$) be the energy density carried by the electromagnetic field (respectively stored in the atoms). The total flux of energy through a section perpendicular to propagation direction is $(U_{em}^{in}+U_{at})v_{g}$. This coincides with the flux of electromagnetic energy, provided the latter quantity is given by $U_{em} c$, as can be easily derived from Eq. \ref{energy_at}. This problem was already addressed in Ref. \cite{courtens1968}, in the framework of self induced transparency~\cite{mccall1967}.

\section{Experiments}
We investigate slow light propagation using spectral hole burning in a thulium doped YAG crystal. This material fulfills two essential conditions for the creation of a spectral hole. First, the inhomogeneous broadening ($\Gamma_{inh}\approx$ 20 GHz) is much larger than the homogeneous linewidth ($\gamma_{hom}\approx$ 10 kHz, at 1.8K). Second, thulium ions exhibit a double lambda structure under a properly oriented external magnetic field. This four level structure consists of two ground, and two excited hyperfine sublevels~\cite{seze2006}: one of the ground states can be used as a shelving state for persistent spectral hole burning .

The system is shined with an extended cavity diode laser (ECDL) operating at 793 nm, stabilized on a high finesse Fabry-Perot cavity through a Pound-Drever-Hall servoloop. The laser line width is reduced to 200 Hz over 10 ms~\cite{crozatier2004}. A semiconductor tapered amplifier (Toptica BoosTA) is used to raise the beam intensity.

We resort to acousto-optic modulators (AOM) for the temporal shaping of the optical field phase and amplitude. Spatial filtering by a single mode fiber precisely controls the spatial phase and amplitude. However, combining temporal and spatial control requires some care. Indeed, the direction of an AOM deflected beam varies with the driving RF frequency. In double-pass configuration, the beam emerges in fixed direction but a single AOM is then not enough to carry over an arbitrary phase and amplitude shaping onto the optical field. For instance, modulation at RF frequencies f$_{1}$ and f$_{2}$ in double pass arrangement gives rise to shifted components, not only at 2f$_{1}$ and 2f$_{2}$, but also at f$_{1}+$f$_{2}$. To get rid of this beat note, one has to use two distincts AOMs, each one being driven at a single RF frequency. On that purpose the laser beam is directed to a polarizing beam-splitter that separates the two orthogonal light polarizations. Each one of the split beams is double-passed through an AA Opto-Electronic acousto-optic modulator centered at 110 MHz. The two AOMs are independently and synchronously driven by a dual-channel 1Gigasample/s waveform generator (Tektronix AWG520) that can provide arbitrary amplitude and phase shaping. Each channel feeds one frequency-shift at a time. After passing twice through the AOMs, the split beams come back to the cube where they merge. The recombined beam is routed in a fixed direction, insensitive to frequency shift. The merged beam is finally coupled into a monomode fiber. Therefore all the optical field components propagate along the same spatial mode whatever their frequency. It is then focused on a 0.5 at.\% Tm${}^{3+}$:YAG sample cooled down to 1.7K in an Oxford Spectromag cryostat. The magnetic field generated by superconducting coils is applied in the direction giving maximum branching ratio~\cite{seze2006}. The spot diameter on the crystal is adjusted to 80$\mu$m. Sample is then imaged onto a 50$\mu$m-diameter pinhole with a magnification factor of 2. The transmitted light is collected on an avalanche photodiode HAMAMATSU C5460 or C4777 protected from strong excitation pulses by a third AOM used as a variable density filter.
\begin{figure}[!ht]
\begin{centering}
\includegraphics[width=8cm]{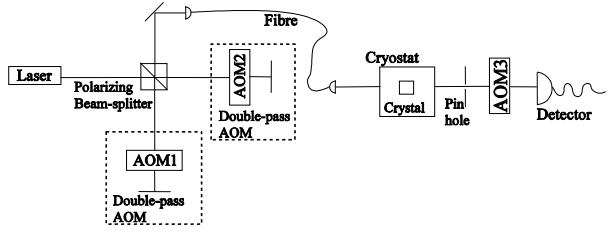}
\caption{Experimental set-up. The acousto-optic modulators AOM1 and 2 are set in double-pass configuration so that the deflected beam direction does not vary with the RF driving frequency. The various beams are all put into the unique spatial mode of a single mode fiber before reaching the liquid helium-cooled crystal . AOM3 is used as a gate to protect the detector from the intense preparation pulses.}\label{exp}
\end{centering}
\end{figure}

\begin{figure}[!ht]
\begin{centering}
\includegraphics[width=8cm]{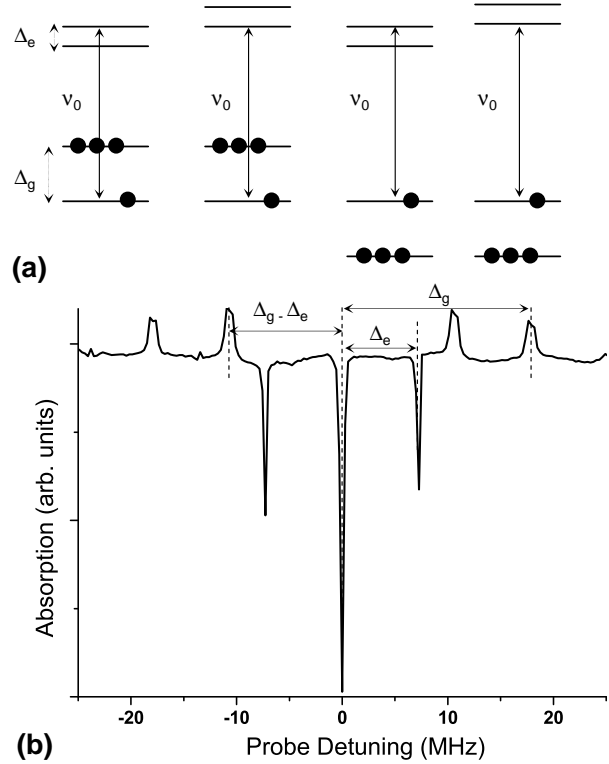}
\caption{Hole burning principle. (a) Simultaneous excitation of four classes of atoms by laser at frequency $\nu_{0}$. Atoms, represented by circles, are pumped in the non-resonant ground sublevel. $\Delta_{g}$ and $\Delta_{e}$ respectively stand for ground and excited state splittings (b) Hole burning absorption spectrum. We have in this case $\Delta_{g} = $18 MHz, and $\Delta_{e} =$ 7.5 MHz.}\label{hole-burning}
\end{centering}
\end{figure}

Persistent spectral hole burning in Tm:YAG is described in Ref.~\cite{louchet2007} and is illustrated in Fig.~\ref{hole-burning}. At 1.8K, both ground states are equally populated. The crystal is illuminated by a laser at frequency $\nu_{0}$. Due to inhomogeneous broadening, the laser simultaneously excites four classes of ions and optical pumping tends to unbalance the ground level distribution, around excitation frequency $\nu_{0}$.
A weak probe pulse absorption depends on the ground state population. Hence, around $\nu_{0}$ and $\nu_{0} \pm \Delta_{e}$, absorption is decreased: there is a transparency window, also called a hole, in the absorption profile. Conversely, around $\nu_{0}\pm \Delta_{g}$ and $\nu_{0} \pm \left( \Delta_{g} - \Delta_{e} \right)$ an anti-hole reflects an absorption increase. In figure~\ref{hole-burning}(b) we show  an absorption spectrum, when a probe field is chirped around excitation frequency $\nu_{0}$. Hence, hole burning provides a simple way to create a transparency window in an inhomogeneous medium. Once the hole is burnt, a signal pulse, whose spectrum is contained within the spectral hole, is sent through the crystal.

\begin{figure}[!ht]
\begin{centering}
\includegraphics[width=8cm]{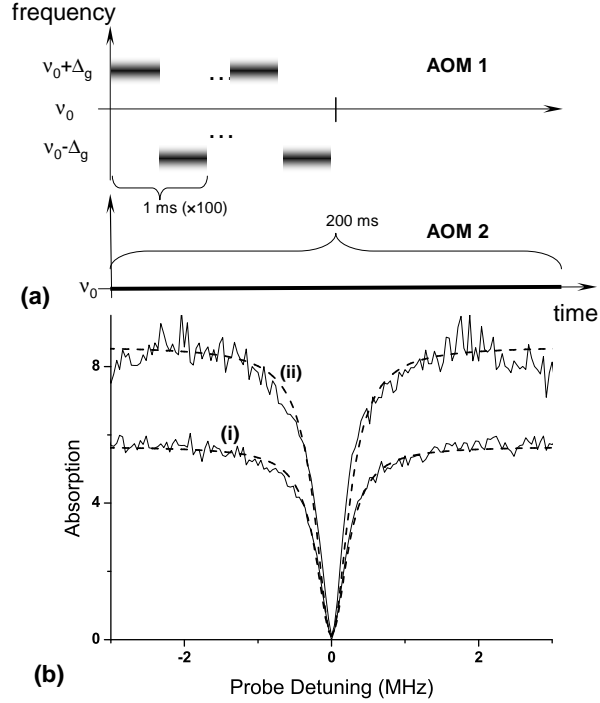}
\caption{(a) time diagram of the hole burning sequence. Three different frequencies are involved in the preparation step. In addition to hole burning at $\nu_{0}$, one increases the available optical depth by pumping around $\nu_{0} + \Delta_{g}$ and $\nu_{0} - \Delta_{g}$. AOM2 is used to burn the hole, while pumping to the hole region is accomplished by alternatively chirping AOM1 around $\nu_{0} + \Delta_{g}$ and $\nu_{0} - \Delta_{g}$. (b) Hole burnt at $\nu_{0}$ with AOM1 off (i) and on (ii). The maximum hole depth is 8.5 cm${}^{-1}$.}\label{preparation}
\end{centering}
\end{figure}

In order to investigate slow light, we aim at creating a TW as deep and narrow as possible in Tm${}^{3+}$:YAG crystal. In our 0.5\% doped crystal, linear absorption coefficient is 5 cm${}^{-1}$. Since both ground sublevels are equally populated before hole burning, one can increase the optical depth around $\nu_{0}$ by pumping the ions around $\nu_{0} + \Delta_{g}$ and $\nu_{0} - \Delta_{g}$. During the preparation step, one simultaneously pumps at the hole burning position $\nu_{0}$ with AOM2, and over a few MHz-wide intervals centered at $\nu_{0} + \Delta_{g}$ and $\nu_{0} - \Delta_{g}$ with AOM1 (see Fig.~\ref{preparation}a). As already mentioned, our double-pass AOMs only manage a single frequency at a time. Therefore one has to chirp AOM1 alternatively around $\nu_{0} + \Delta_{g}$ and $\nu_{0} - \Delta_{g}$. Compound generation of both frequencies would give rise to spurious excitation around the beat-note $\nu_{0}$. With this procedure the available absorption is raised up to 8.5 cm${}^{-1}$. The linear absorption coefficient around the spectral hole, with and without the absorption improvement sequence, is displayed in Fig.~\ref{preparation}(b). In both cases, the hole is well described by a lorentzian shape.

\begin{figure}[!ht]
\begin{centering}
\includegraphics[width=8cm]{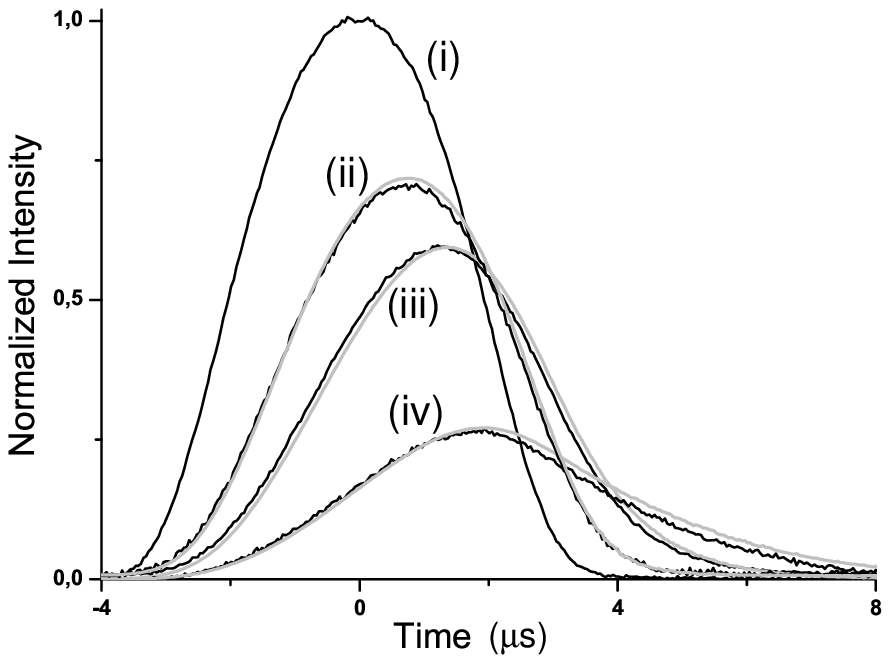}
\caption{Pulse propagation through several spectral holes. The input pulse coincides with the reference pulse (i). Transmission through holes of width 206 kHz, 420 kHz and 860 kHz respectively leads to the delayed, attenuated and stretched profiles numbered from (ii) to (iv). The hole depth is maintained at 8.5 cm${}^{-1}$. The theoretical profiles are completely determined by the hole width and depth. They are displayed as grey lines, together with the experimental data, represented by black lines.}\label{slow_down}
\end{centering}
\end{figure}

Once preparation is completed, we investigate the propagation of a weak signal, with a maximum power of $\approx 2\mu$W, through the spectral hole burnt at frequency $\nu_{0}$. Figure~\ref{slow_down} represents the transmission of $5.37\mu$-long pulses through a spectral hole, for various hole-width values. One varies the hole width from 206 kHz to 860 kHz by adjusting the pumping power from 100$ \mu$W to 300$ \mu$W. 
The origin of time and the input signal shape are given by a reference pulse that propagates through a 100 \% transmission, 10 MHz-wide spectral hole. For 860 and 420 kHz-wide holes, the pulse is not distorded, and the delay is simply deduced from the maximum amplitude position. For the 200 kHz-wide hole, we measure a delay of 2 $\mu$s, which corresponds to a group velocity of 2500 ms${}^{-1}$.

\begin{figure}[!ht]
\begin{centering}
\includegraphics[width=8cm]{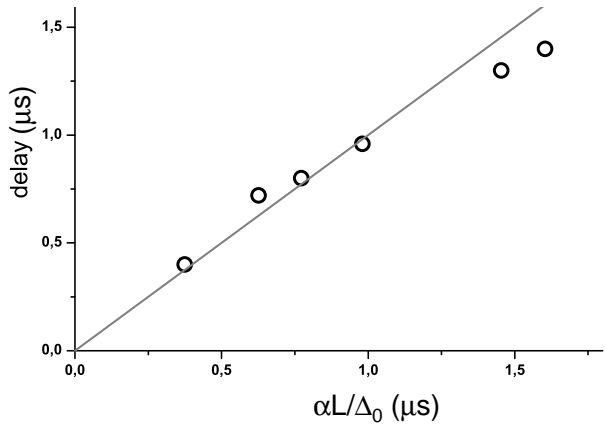}
\caption{experimentally measured delays as a function of the measured ratio $\alpha L/\Gamma$. The delay caused by a lorentzian hole is expected to equal this quantity.}\label{delays}
\end{centering}
\end{figure}

For a lorentzian spectral hole, the delay experienced by a gaussian pulse is simply expressed as $\alpha L/\Gamma$. In order to check this dependence, we measure the delay of the 5.37$ \mu s$-long gaussian pulses as a function of the ratio $\alpha L/\Gamma$. To prevent pulse distortion, we keep the hole width larger than 600 kHz. We control $\alpha L/\Gamma$ by varying the hole depth. The measured delay is plotted as a function of the ratio $\alpha L/\Gamma$ in Fig.~\ref{delays}. The experimental data agree well with theoretical prediction.

\section{Conclusion}
For the first time in this kind of material, we have observed the slowing down of light produced by persistent spectral hole burning in a thulium doped YAG crystal. The largest value of the observed delay, 2 $\mu$s, corresponds to a group velocity reduction by a factor of more than $10^5$. A more concentrated and longer crystal may provide larger delays. Indeed, the upper limit for the fractional delay is limited by the opacity $\alpha L$. The hole is burnt by optical pumping to an other hyperfine sub-level. The long hyperfine state lifetime entails important consequences. On the one hand, taking advantage of the hole persistence, we were able to devise a sophisticated pumping scheme that enabled us to significantly increase the avalaible opacity, which has resulted in an improved hole depth. On the other hand, the hole can be prepared long before the slow light observation, which eliminates the simultaneous presence of an intense control field. Analyzing the slowing down process, we have pointed out the role played by off resonance atoms where most of the incoming information is carried over while the pulse is confined within the sample. This might open a way to quantum storage, quite similarly to EIT.


\begin{thebibliography}{1}
\bibitem{brillouin1960} L. Brillouin, Wave Propagation and Group Velocity (Academic, New York, 1960).
\bibitem{garrett1970} G. G. B. Garrett, and D. E. McCumber, Phys. Rev. A {\bf 1}, 305 (1970).
\bibitem{chu1982} S. Chu, and S. Wong, Phys. Rev. Lett. {\bf 48}, 738 (1982).
\bibitem{wang2000} L. J. Wang, A. Kuzmich, and A. Dogariu, Nature {\bf 406}, 277 (2000).
\bibitem{harris1990} S. E. Harris, J. E. Field, and A. Imamoglu, Phys. Rev. Lett. {\bf 64}, 1107 (1990).
\bibitem{camacho2007} R. M. Camacho, C. J. Broadbent, I. Ali-Khan, and J. C. Howell, Phys. Rev. Lett. {\bf 98}, 043902 (2007).
\bibitem{lukin2003} M. D. Lukin, Rev. Mod. Phys. {\bf 75}, 457 (2003).
\bibitem{hau1999} L. V. Hau, S. E. Harris, Z. Dutton, and C. H. Behroozi, Nature(London) {\bf 397}, 594 (1999).
\bibitem{kasapi1995} A. Kasapi, M. Jain, G.Y. Yin, and S. E. Harris, Phys. Rev. Lett. {\bf 74}, 2447 (1995).
\bibitem{wei1999} C. Wei, and N. B. Manson, Phys. Rev. A {\bf 60}, 2540 (1999).  
\bibitem{budker2000} D. Budker, D. F. Kimball, S. M. Rochester, and V. V. Yashchuk, Phys. Rev. Lett. {\bf 83}, 1767 (1999).
\bibitem{turukhin2002} A. V. Turukhin, V. S. Sudarshanam, M. S. Shahriar, J. A. Musser, B. S. Ham, and P. R. Hemmer, Phys. Rev. Lett. {\bf 88}, 023602 (2001).
\bibitem{bigelow2003} M. S. Bigelow, N. N. Lepeshkin, and R. W. Boyd, Phys. Rev. Lett. {\bf 90}, 113903 (2003).
\bibitem{longdell2005} J. J. Longdell, E. Fraval, M. J. Sellars, and N. B. Manson, Phys. Rev. Lett. {\bf 95}, 063601 (2005).
\bibitem{baldit2005} E. Baldit, K. Bencheikh, P. Monnier, J. A. Levenson, and V. Rouget, Phys. Rev. Lett. {\bf 95}, 143601 (2005).
\bibitem{harris1997} S. E. Harris, Physics Today {\bf 50(7)}, 36 (1997).
\bibitem{shwarz1967} S. E. Shwarz, and T. Y. Tan, Appl. Phys. Lett. {\bf 10}, 4 (1967).
\bibitem{camacho2006} R. M. Camacho, M. V. Pack, and J. C. Howell, Phys. Rev. A {\bf 74}, 033801 (2006).
\bibitem{shakhmuratov2005} R. N. Shakhmuratov, A. Rebane, P. M\'egret, and J. Odeurs, Phys. Rev. A. {\bf 71}, 053811 (2005).
\bibitem{rebane2007} A. Rebane, R. N. Shakhmuratov,  P. M\'egret, and J. Odeurs, J. Lum. {\bf 127}, 22 (2007).
\bibitem{konz2003} F. K\"onz, Y. Sun, C. W. Thiel, R. L. Cone, R. W. Equall, R. L. Hutcheson, and R. M. Macfarlane, Phys. Rev. B {\bf 68}, 085109 (2003).
\bibitem{grischkowsky1973} D. Grischkowsky, Phys. Rev. A. {\bf 7}, 2096 (1973).
\bibitem{courtens1968} E. Courtens, Phys. Rev. Lett. {\bf 21}, 3 (1968).
\bibitem{mccall1967} S. L. McCall, and E. L. Hahn, Phys. Rev. Lett. {\bf 18}, 908 (1967).
\bibitem{seze2006} F. de S\`eze, A. Louchet, V. Crozatier, I. Lorger\'e, F. Bretenaker, J.-L. Le Gou\"et, O. Guillot-No\"el and Ph. Goldner, Phys. Rev. B {\bf 73}, 085112 (2006).
\bibitem{louchet2007} A. Louchet, J. S. Habib, V. Crozatier, I. Lorger\'e, F. Goldfarb, F. Bretenaker, J.-L. Le Gou\"et, O. Guillot-No\"el and Ph. Goldner, Phys. Rev. B {\bf 75}, 035131 (2007).
\bibitem{crozatier2004} V. Crozatier, F. de S\`eze, L. Haals, F. Bretenaker, I. Lorger\'e, and J.-L. Le Gou\"et, Opt. Commun. {\bf 241}, 203 (2004).
\end{thebibliography}
\end{document}